\documentclass[preprint,showpacs,preprintnumbers,amsmath,amssymb]{revtex4}

\begin{document}


\title{Cosmological versions of Vaidya's radiating stellar exterior,
an accelerating reference frame, and Kinnersley's photon rocket}

\author{M. L. McClure, Kaem Anderson, and Kirk Bardahl}
\email{mcclure@astro.utoronto.ca}
\affiliation{Mathematics \& Applied Mathematics, University of Cape Town,
Rondebosch 7701, South Africa}

\date{September 21, 2008}

\begin{abstract}

The spacetimes for Vaidya's radiating stellar exterior and Kinnersley's 
photon rocket in a cosmological background are obtained by performing 
the same conformal transformation as is used to obtain the 
Robertson-Walker metric from Minkowski spacetime.
In the case of the cosmological radiating stellar exterior, a two-fluid 
solution is found that obeys all of the energy conditions and consists 
of a null fluid and a perfect fluid that asymptotically falls off to 
the standard cosmological values for pressure and density at infinite radius. 
For the cosmological photon rocket, the massless case is first interpreted 
to obtain a solution for an accelerating cosmological reference frame, 
and then the general case is interpreted: in both cases, a two-fluid  
solution is found that consists of a null fluid and an imperfect fluid 
that possesses heat conduction and anisotropic stress.
The imperfect fluid appears to contain an inhomogeneous dark energy 
component that acts to accelerate the matter through space via a pressure 
gradient, but this component has negative energy density on the trailing 
side of the rocket, meaning only the leading side of the rocket is guaranteed 
to satisfy the weak and dominant energy conditions.
Unlike spacetimes that have rotation but no acceleration, the cosmological
photon rocket can serve as an example of a spacetime that contradicts Mach's 
notion of acceleration, since an observer would see empirical evidence of 
acceleration even though the matter does not accelerate relative to the 
universe's background matter distribution.

\end{abstract}

\pacs{04.20.Jb, 04.40.Nr, 04.70.Bw, 98.80.Jk}

\maketitle

\section{Introduction}

A time-dependent conformal transformation of Minkowski space can yield the
Robertson-Walker metric representing expanding matter-filled Friedmann
universes.  This is because the conformal transformation makes
the spacetime dynamic and introduces mass-energy, which is consistent
since a homogeneous matter content can cause an acceleration in the
contraction of space backward-in-time toward the Big Bang, or equivalently
cause a deceleration in the expansion of space forward-in-time, due to
gravity.  This same conformal transformation will be used in this paper on
spacetimes related to Minkowski space (via a Kerr-Schild metric
transformation \cite{Ker65}) with the goal of obtaining solutions in a
cosmological background, ideally consisting of incoherent radiation or
dust. These solutions should be less simplistic than the original
solutions and less simplistic than homogeneous cosmological models,
creating exact solutions of Einstein's field equations of greater
sophistication. 

Thakurta \cite{Tha81} performed a conformal transformation of the 
Schwarzschild metric to obtain a spacetime that locally looks like 
Schwarzschild and asymptotically looks like Robertson-Walker at infinite 
radius; however the energy-momentum tensor was not fully interpreted 
(Krasi\'{n}ski \cite{Kra97}), and it is known at least for the case 
of an asymptotically flat universe that it violates energy conditions 
within the event horizon (McClure and Dyer \cite{McC06}). 
Sultana and Dyer \cite{Sul05} performed a conformal transformation of the
Eddington-Finkelstein form (which is also a Kerr-Schild \cite{Ker65} 
form) of the Schwarzschild metric, yielding a different energy-momentum 
tensor that is unphysical within a radius that grows to infinity at 
infinite time; however, it can be shown that the white hole form yields a 
physical solution throughout the spacetime (as was calculated 
inadvertently by Sultana \cite{Sul03}). 
Thus, conformal transformations of spacetimes like the Eddington-Finkelstein 
white hole should be promising as potential cosmological solutions.

Vaidya \cite{Vai43} devised a metric for a radiating white hole by using 
the Schwarzschild solution and allowing the singularity mass to vary as a 
function of $u=t-r$ such that the white hole mass is radiated away on 
null surfaces. 
The metric was presented in Kerr-Schild form for the radiating Kerr white 
hole (Vaidya and Patel \cite{Vai73}).
Kinnersley \cite{Kin69} devised a solution for a white hole that 
accelerates by radiating anisotropically, so it is the accelerated 
form of Vaidya's radiating stellar exterior.
The photon rocket was presented in more detail and also extended to 
the charged case by Kinnersley and Walker \cite{Kin70}.  

In this paper solutions will be sought for the Kerr-Schild forms of
Vaidya's radiating stellar exterior and Kinnersley's photon rocket in a
cosmological background. 
Since the Kerr-Schild form of a white hole yields a solution in 
cosmological background \cite{Sul03}, it can be expected that conformal 
transformations of the Kerr-Schild forms of the Vaidya and Kinnersley 
metrics may yield cosmological solutions. 
The calculations in this paper will be performed using the \textsc{Redten} 
package (Harper and Dyer \cite{Har94}) and the computer algebra program 
\textsc{Reduce}. 

It should be noted that previously Vaidya and Patel \cite{Vai89} studied the
Kerr metric in the Einstein static universe, which they assumed to be a
mixture of a null fluid and perfect fluid, although they do not appear to have
verified that this interpretation obeys the energy conditions.
Obtaining solutions for cosmological Kerr black holes is problematic, 
since they must necessarily frame-drag the surrounding universe, so 
finding a solution for the exterior of a non-isolated Kerr black hole is 
somewhat like trying to find an exact solution for a Kerr interior. 
Thus, simply finding a solution of a spherically-symmetric radiating stellar 
exterior in an expanding universe should be more feasible and more 
realistic of our expanding Universe, so Kerr black holes will not
be considered in this paper. 

While it is expected that the conformal transformation should introduce a
matter distribution similar to that of an Friedmann universe, an interesting
problem is whether the photon rocket will accelerate with respect to the
matter or whether it is possible for the entire matter distribution to
accelerate along with the rocket.  If the matter distribution does not
accelerate with the rocket, it must avoid entering the event horizon of
the photon rocket since it is a white hole.  Also, the photon rocket has
no gravitational radiation (Bonnor \cite{Bon94}) and the conformal
transformation must preserve the causal structure such that there is no
gravitational radiation despite the presence of the universe's background
matter.  Thus, finding a solution in which the background matter does not
accelerate with the rocket seems more complicated than finding a solution
in which it does. 

If the matter distribution does accelerate with the rocket, it would 
be an unusual universe, but the existence of such a solution would be 
interesting as an example of a universe in which an object can have
absolute acceleration without any acceleration relative to
the universe's background matter distribution.  This would be in
contradiction to Mach's notion that acceleration is relative.
While solutions with rotation such as the dust solution of Lanczos
(\cite{Lan24}) have demonstrated that it is possible to have a universe with
rotation, it should be noted that beyond Newtonian physics, rotation is
not sufficient for the existence of acceleration.  Since gravitational
forces are geometrized in general relativity, matter that simply travels
inertially along geodesics to undergo rotation actually has no force acting 
on it or accelerating it. Thus, to truly show that general relativity defies 
Mach's notion of acceleration, it is necessary to explore spacetimes in 
which there is actual acceleration of the matter, meaning spacetimes in
which there are non-gravitational forces. 

\section{Preliminaries}

Kerr-Schild metrics \cite{Ker65} can be expressed as
\begin{equation}
\bar{g}_{ab} = g_{ab} + 2H l_a l_b
,
\end{equation}
where $g_{ab}$ is Minkowski spacetime (or in the generalized case, any seed 
metric), $2H$ is a scalar field, and $l^a$ is a null vector 
field of the seed metric (and the transformed metric).
The Schwarzschild metric can be expressed in Kerr-Schild form as
\begin{equation}
\mathrm{d}s^2 = - \mathrm{d}{t}^2 + \mathrm{d}r^2 + r^2
                \left(\mathrm{d}{\theta}^2 + \mathrm{sin}^2 \theta \; 
                \mathrm{d} {\phi}^2 \right) + \frac{2m}{r}(\mathrm{d}t \pm 
                \mathrm{d}r)^2 
,
\end{equation}
with Minkowski spacetime as the seed metric, $2H=2m/r$ as the scalar 
field, and $(1,\pm1,0,0)$ as the null vector field.
The plus or minus sign corresponds to the black hole or white hole case 
respectively (since it determines the sign of the $\mathrm{d}t 
\mathrm{d}r$ cross term in the metric so that the white hole case is the 
temporal reverse of the black hole case).
This metric is related to the standard form of the Schwarzschild metric 
by a co-ordinate transformation and is equivalent to the 
Eddington-Finkelstein form of the Schwarzschild metric.

Vaidya's radiating stellar exterior \cite{Vai43} can be expressed in 
Kerr-Schild form as 
\begin{equation}
\mathrm{d}s^2 = - \mathrm{d}{t}^2 + \mathrm{d}r^2 + r^2 
                \left(\mathrm{d}{\theta}^2 + \mathrm{sin}^2 \theta \; 
                \mathrm{d} {\phi}^2 \right) + 
                \frac{2m(t,r)}{r}(\mathrm{d}t-\mathrm{d}r)^2
, 
\end{equation}
where the white hole mass $m$ varies across null surfaces as a function of 
$u=t-r$ such that 
\begin{equation}
\frac{\mathrm{d}m}{\mathrm{d}t} = - \frac{\mathrm{d}m}{\mathrm{d}r} = m' 
\end{equation}
as the white hole mass is radiated away.
Looking at the Einstein tensor---with $G_{ab}=-\kappa T_{ab}$ 
(negative Einstein sign convention)---the only non-zero components are
\begin{equation}
G^0_0 = G^1_0 = - G^0_1 = -G^1_1 = -\frac{2m'}{r^2} 
,
\end{equation}
which is interpreted as a null fluid.
The energy-momentum tensor can be written as $T^{ab} = \tau l^a l^b$ with 
$l^a l_a = 0$ such that $G^a_a=0$.

Kinnersley's photon rocket \cite{Kin69} with uniform acceleration $a$ is
\[
\mathrm{d}s^2 = - \mathrm{d}{t}^2 + \mathrm{d}r^2 + r^2 
                \left(\mathrm{d}{\theta}^2 + \mathrm{sin}^2 \theta \; 
                \mathrm{d} {\phi}^2 \right) + 2 a r^2 \mathrm{sin} \, 
                \theta \; \mathrm{d}\theta (\mathrm{d}t-\mathrm{d}r) 
\]
\begin{equation}
+\left(a^2 r^2 \mathrm{sin}^2 \theta + 2 a r \, \mathrm{cos} \, \theta + 
                   \frac{2m(t,r)}{r} \right) (\mathrm{d}t-\mathrm{d}r)^2
.
\end{equation}
In general the magnitude of the acceleration can vary and it can change 
direction, but the acceleration will be assumed to be uniform for the 
purposes of this paper. 
Looking at the Einstein tensor, the only non-zero components are 
\begin{equation}
G^0_0 = G^1_0 = - G^0_1 = -G^1_1 =  \frac{2}{r^2} (3am \, \mathrm{cos} \, 
\theta - m')
, 
\end{equation}
which can be interpreted as a null fluid that radiates anisotropically to 
accelerate the white hole.

Sultana \cite{Sul03} obtained a solution for a white hole in Einstein-de 
Sitter dust (although it was referred to as a black hole solution)
by performing a conformal transformation on the 
Eddington-Finkelstein form of the Schwarzschild metric with conformal 
factor $\Omega^2 = t^4$.
Generalizing this cosmological white hole for a scale factor $R(t)$ for 
any flat Robertson-Walker metric, this spacetime can be expressed as 
\begin{equation}
\mathrm{d}s^2 = [R(t)]^2 \left( - \mathrm{d}{t}^2 + \mathrm{d}r^2 + r^2
                \left(\mathrm{d}{\theta}^2 + \mathrm{sin}^2 \theta \; 
                \mathrm{d} {\phi}^2 \right) + \frac{2m}{r}(\mathrm{d}t - 
                \mathrm{d}r)^2 \right)
,
\end{equation}
where the time co-ordinate is related to the usual cosmological time 
co-ordinate---here designated as $t_c$---by a co-ordinate transformation 
$R(t)\mathrm{d}t = \mathrm{d}t_c$.
For a flat radiation universe, the scale factor 
evolves compared with some time $t_o$ at which $R=1$ as $R(t_c) = 
(t_c/t_{c,o})^{1/2}$, which is equivalent to $R(t) = t/t_o$.
For a flat dust universe the scale factor evolves 
as $R(t_c) = (t_c/t_{c,o})^{2/3}$, which is equivalent to $R(t) = (t/t_o)^2$.

The non-zero components of the Einstein tensor are
\begin{equation}
G^0_0 = \frac{3\dot{R}^2}{R^4}\left(1+\frac{2m}{r}\right) +
        \frac{4\dot{R}m}{R^3r^2}
\end{equation}
\begin{equation}
G^0_1 = \frac{2\dot{R}m}{R^3r^2}
\end{equation}
\begin{equation}
G^1_0 = \left(\frac{2\dot{R}^2}{R^4} - \frac{\ddot{R}}{R^3}\right)
         \left(\frac{4m}{r}\right) - \frac{2\dot{R}m}{R^3r^2}
\end{equation}
\begin{equation}
G^1_1 = -\left(\frac{\dot{R}^2}{R^4} - \frac{2\ddot{R}}{R^3}\right)
        \left(1+\frac{2m}{r}\right) + \frac{8\dot{R}m}{R^3r^2}
\end{equation}
\begin{equation}
G^2_2 = G^3_3 = -\left(\frac{\dot{R}^2}{R^4} - \frac{2\ddot{R}}{R^3}\right)
                \left(1+\frac{2m}{r}\right) 
.
\end{equation}
The energy-momentum tensor can be interpreted as a two-fluid 
solution consisting of a null fluid and a perfect fluid.
The null fluid is given by
\begin{equation}
{\tau l^0 l_0} = {\tau l^1 l_0} = - {\tau l^0 l_1} = -{\tau l^1 l_1} = 
                 \frac{2\dot{R} m}{\kappa R^3 r^2}
.
\end{equation}
The perfect fluid is given by
\begin{equation}
T^a_b = (\mu +p)u^a u_b + p g^a_b
.
\end{equation}
From spherical symmetry the angular components of the 
velocity field must be zero ($u^2=u^3=0$), so the isotropic pressure is 
given from $G^2_2 = G^3_3 = - \kappa p$ as
\begin{equation}
p = \frac{1}{\kappa}\left(\frac{\dot{R}^2}{R^4}-\frac{2\ddot{R}}{R^3}\right)
    \left(1+\frac{2m}{r}\right)
.
\end{equation}  
The energy density $\mu$ is determined from the trace of the Einstein 
tensor ($G^a_a = \kappa(\mu-3p)$), which yields
\begin{equation}
\mu = \frac{3\dot{R}^2}{\kappa R^4}\left(1+\frac{2m}{r}\right)+
      \frac{12\dot{R}m}{\kappa R^3r^2}
.
\end{equation}
To satisfy the energy conditions (e.g.\ see Wald \cite{Wal84}), the weak 
energy condition requires that $\mu \ge 0$ and $\mu+p_i \ge 0$, the strong 
energy condition requires that $\mu + \Sigma p_i \ge 0$ and $\mu+p_i \ge 
0$, and the dominant energy condition requires that $\mu \ge |p_i|$.
For an expanding universe, $\dot{R}/R > 0$, the energy density is 
clearly positive everywhere for the cosmological white hole. For $R(t)$ 
that goes as $t^x$ with $0 < x < 2$, the pressure is positive.
Neglecting the last term of $\mu$ (which dominates at late times and is 
insignificant at small times compared with 
the pressure and the first part of the expression for $\mu$) the pressure 
is $p >> \mu$ near $x=0$, $p = \mu$ at $x = 1/2$, $p = \mu/3$ at 
$x=1$, and $p=0$ at $x=2$. For $x > 2$ the pressure is negative, only 
asymptotically approaching $p \le -\mu/3$.
Since the last term of $\mu$ only serves to make $\mu$ larger for an 
expanding universe, then the 
pressure can only be less than or equal to the forementioned relations.
Thus, clearly all the energy conditions are satisfied by the cosmological 
white hole at all times if the universe is expanding with $x > 1/2$ (which 
is in contrast to the corresponding cosmological black hole of Sultana 
and Dyer \cite{Sul05}).

\section{A cosmological radiating stellar exterior}

A conformal transformation can be performed on Vaidya's metric to obtain a 
metric that is asymptotically Roberston-Walker as $r$ goes to 
infinity.
For flat Robertson-Walker this yields
\begin{equation}
\mathrm{d}s^2 = [R(t)]^2 \left( - \mathrm{d}{t}^2 + \mathrm{d}r^2 + 
r^2 \left(\mathrm{d}{\theta}^2 + \mathrm{sin}^2 \theta \; \mathrm{d} {\phi}^2 
\right) + \frac{2m(t,r)}{r} (\mathrm{d}t-\mathrm{d}r)^2 \right) , 
\end{equation}
where the scale factor $R(t)$ depends on $t$ only.
The non-zero Einstein tensor components of this metric are
\begin{equation}
G^0_0 = \frac{3\dot{R}^2}{R^4}\left(1+\frac{2m}{r}\right)+
\frac{2\dot{R}}{R^3r}\left(\frac{2m}{r}-m'\right)-\frac{2m'}{R^2r^2}
\end{equation}
\begin{equation}
G^0_1 = \frac{2\dot{R}}{R^3r}\left(\frac{m}{r}+m'\right)+\frac{2m'}{R^2r^2}
\end{equation}
\begin{equation}
G^1_0 = \left(\frac{2\dot{R}^2}{R^4}-\frac{\ddot{R}}{R^3}\right)
         \left(\frac{4m}{r}\right)
        -\frac{2\dot{R}}{R^3r}\left(\frac{m}{r}+m'\right)-\frac{2m'}{R^2r^2}
\end{equation}
\begin{equation}
G^1_1 = -\left(\frac{\dot{R}^2}{R^4}-\frac{2\ddot{R}}{R^3}\right)
        \left(1+\frac{2m}{r}\right)
        +\frac{2\dot{R}}{R^3r}\left(\frac{4m}{r}+m'\right)+\frac{2m'}{R^2r^2}
\end{equation}
\begin{equation}
G^2_2 = G^3_3 = -\left(\frac{\dot{R}^2}{R^4}-\frac{2\ddot{R}}{R^3}\right)
                \left(1+\frac{2m}{r}\right) .
\end{equation}
In the case where $m' = 0$, this reduces to the white-hole version of 
Sultana and Dyer's \cite{Sul05} cosmological black hole.
The $m'$ terms in $G^0_0$, $G^0_1$, $G^1_0$, and $G^1_1$ can 
be interpreted as an additional component of the null fluid due to the 
radiation of the point mass, analogous to Vaidya's radiating star.
Thus, assuming a solution that consists of a superposition of a 
null fluid and a perfect fluid, then the null fluid component of the 
energy-momentum tensor is 
\begin{equation}
{\tau l^0 l_0} = {\tau l^1 l_0} = - {\tau l^0 l_1} = -{\tau l^1 l_1} = 
\frac{2\dot{R} m}{\kappa R^3 r^2}
+ \frac{2m'\dot{R}}{\kappa R^3 r}
+ \frac{2m'}{\kappa R^2 r^2}
.
\end{equation}

Looking at the remaining terms of the Einstein tensor with the null fluid 
component ${G^a_b}_{(nf)}$ subtracted reveals
\begin{equation}
G^0_0 - {G^0_0}_{(nf)} = \frac{3\dot{R}^2}{R^4}\left(1+\frac{2m}{r}\right)+
\frac{6\dot{R}m}{R^3r^2}
\end{equation}
\begin{equation}
G^0_1 - {G^0_1}_{(nf)} = 0
\end{equation}
\begin{equation}
G^1_0 - {G^1_0}_{(nf)} = 
         \left(\frac{2\dot{R}^2}{R^4}-\frac{\ddot{R}}{R^3}\right)
         \left(\frac{4m}{r}\right)
\end{equation}
\begin{equation}
G^1_1 - {G^1_1}_{(nf)} = 
        -\left(\frac{\dot{R}^2}{R^4}-\frac{2\ddot{R}}{R^3}\right)
        \left(1+\frac{2m}{r}\right)
        +\frac{6\dot{R}m}{R^3r^2}
\end{equation}
The remaining heat conduction component is due to the radial 
velocity field component $u^1$ of the perfect fluid, which leads to an 
effective heat conduction due to the flow of matter relative to the $r$ 
co-ordinate.
Spherical symmetry implies there can only be a radial component of the 
velocity field ($u^2=u^3=0$), so $G^2_2 = G^3_3 = - \kappa p$ yields the 
isotropic pressure to be
\begin{equation}
p = \frac{1}{\kappa}\left(\frac{\dot{R}^2}{R^4}-\frac{2\ddot{R}}{R^3}\right)
    \left(1+\frac{2m}{r}\right)
.
\end{equation}
Since $G^a_a=0$ for the null fluid component, then the energy density of 
the perfect fluid is determined from 
\begin{equation}
G^a_a = \kappa (\mu - 3p) = \frac{6\dot{R}^2}{R^4}\left(1+\frac{2m}{r}\right)+
                            \frac{12\dot{R}m}{R^3r^2}
\end{equation}
to be
\begin{equation}
\mu = \frac{3\dot{R}^2}{\kappa R^4}\left(1+\frac{2m}{r}\right)+
      \frac{12\dot{R}m}{\kappa R^3r^2}.
\end{equation}
The pressure and density of the perfect fluid are identical to that of 
the cosmological white hole in the previous section.
Thus, just like that spacetime, this spacetime must represent a physical 
solution for a scale factor $R(t) = (t/t_o)^x$ with $x \ge 1/2$.

In the case of the scale factor for a radiation universe, $R(t) = t/t_o$, 
this yields 
\begin{equation}
p = \frac{1}{\kappa R^2 t^2} \left(1+\frac{2m}{r}\right)
\end{equation}
and
\begin{equation}
\mu = \frac{3}{\kappa R^2 t^2} \left(1+\frac{2m}{r}\right)+
      \frac{12m}{\kappa R^2 t r^2}
,
\end{equation}
so clearly $\mu$ and $p$ are both positive and $p \le \mu/3$, with the energy density and 
pressure both asymptotically approaching the standard radiation universe 
pressure and density as $r$ goes to infinity.
The equation of state is
\begin{equation}
p = \frac{\mu}{3}-\frac{4m}{\kappa R^2 t r^2}
,
\end{equation}
which is interpreted most simply if the energy density consists of a 
radiation component $\mu_r$ given by
\begin{equation}
\mu_r = \frac{3}{\kappa R^2 t^2} \left(1+\frac{2m}{r}\right)
\end{equation}
and a dust component $\mu_d$ given by
\begin{equation}
\mu_d = \frac{12m}{\kappa R^2 t r^2}
,
\end{equation}
so that the pressure corresponds to the radiation component and the dust 
component is pressureless.

In the case of the scale factor for a dust universe, $R(t) = (t/t_o)^2$, 
this yields $p=0$ and
\begin{equation}
\mu = \frac{12}{\kappa R^2 t^2} \left(1+\frac{2m}{r}\right)+
      \frac{24m}{\kappa R^2 t r^2}
,
\end{equation}
with the density again positive and falling off toward the standard 
dust universe density as $r$ goes to infinity.
Since $p=0$, clearly this equation of state corresponds to pressureless dust.

It is assumed the terms of the Einstein tensor not corresponding to 
the null fluid can be represented by a perfect fluid with $u^0$ 
and $u^1$ components.
If this is so, then 
\begin{equation}
- \kappa [(\mu+p) u^0 u_0 + p] = 
\frac{3\dot{R}^2}{R^4}\left(1+\frac{2m}{r}\right)+
\frac{6\dot{R}m}{R^3r^2}
\end{equation}
\begin{equation}
- \kappa [(\mu+p) u^1 u_1 + p] = 
        -\left(\frac{\dot{R}^2}{R^4}-\frac{2\ddot{R}}{R^3}\right)
        \left(1+\frac{2m}{r}\right)
        +\frac{6\dot{R}m}{R^3r^2}
\end{equation}
with the $T^1_0$ component of the energy-momentum tensor merely 
being due to the $u^1$ component of the velocity field of the perfect 
fluid: 
\begin{equation}
- \kappa [(\mu+p) u^1 u_0] = 
         \left(\frac{\dot{2R}^2}{R^4}-\frac{\ddot{R}}{R^3}\right)
         \left(\frac{4m}{r}\right)
.
\end{equation}
Using the interpreted expressions for $\mu$ (31) and $p$ (29) implies that
\begin{equation}
u^0 u_0 = 
- \frac{\left(\frac{4\dot{R}^2}{R^4}-\frac{2\ddot{R}}{R^3}\right)
        \left(1+\frac{2m}{r}\right)+\frac{6\dot{R}m}{R^3r^2}}
{\left(\frac{4\dot{R}^2}{R^4}-\frac{2\ddot{R}}{R^3}\right)
        \left(1+\frac{2m}{r}\right)
        +\frac{12\dot{R}m}{R^3r^2}}
\end{equation}
and
\begin{equation}
u^1 u_1 = 
- \frac{\frac{6\dot{R}m}{R^3r^2}}
{\left(\frac{4\dot{R}^2}{R^4}-\frac{2\ddot{R}}{R^3}\right)
        \left(1+\frac{2m}{r}\right)
        +\frac{12\dot{R}m}{R^3r^2}}
,
\end{equation}
verifying that
\begin{equation}
u^0 u_0 + u^1 u_1 = -1 = u^a u_a
\end{equation}
as expected for a perfect fluid with a radial velocity field and no 
angular velocity field.

\section{An accelerating cosmological reference frame}

Taking Kinnersley's photon rocket and setting the white hole mass to zero 
yields 
\[
\mathrm{d}s^2 = - \mathrm{d}{t}^2 + \mathrm{d}r^2 + r^2 
\left(\mathrm{d}{\theta}^2 + \mathrm{sin}^2 \theta \; \mathrm{d} {\phi}^2 
\right) \]
\begin{equation}
+ 2 a r^2 \mathrm{sin} \, \theta \; \mathrm{d}\theta 
(\mathrm{d}t-\mathrm{d}r) + \left(a^2 r^2 \mathrm{sin}^2 \theta + 2 a r 
\, \mathrm{cos} \, \theta \right) (\mathrm{d}t-\mathrm{d}r)^2, 
\end{equation}
which is the metric for empty Minkowski space as seen from the reference 
frame of an accelerating observer at $r=0$.
Performing a conformal transformation on this accelerated version of Minkowski 
space yields
\[
\mathrm{d}s^2 = [R(t)]^2 \Big( - \mathrm{d}{t}^2 + \mathrm{d}r^2 + r^2 
\left(\mathrm{d}{\theta}^2 
+ \mathrm{sin}^2 \theta \; \mathrm{d} {\phi}^2 \right)
\]
\begin{equation}
+ 2 a r^2 \mathrm{sin} \, \theta \; \mathrm{d}\theta (\mathrm{d}t-\mathrm{d}r) +
\left(a^2 r^2 \mathrm{sin}^2 \theta + 2 a r \, \mathrm{cos} \, \theta 
\right)(\mathrm{d}t-\mathrm{d}r)^2 \Big) 
.
\end{equation}

Looking at the Einstein tensor, the non-zero components are
\begin{equation}
G^0_0 = \frac{3\dot{R}^2}{R^4} (1 + 2 a r \, \mathrm{cos} \, \theta) + 
        \frac{8\dot{R}}{R^3} (a \, \mathrm{cos} \, \theta) 
\end{equation}
\begin{equation}
G^0_1 = - \frac{2\dot{R}}{R^3} (a \, \mathrm{cos} \, \theta) 
\end{equation}
\begin{equation}
G^1_0 = \left( \frac{2\dot{R}^2}{R^4} - \frac{\ddot{R}}{R^3} \right) (4a 
        r \, \mathrm{cos} \, \theta) + \frac{2\dot{R}}{R^3} (a \, 
        \mathrm{cos} \, \theta) 
\end{equation}
\begin{equation}
G^0_2 = \frac{2\dot{R}}{R^3} (a r \, \mathrm{sin} \, \theta)
\end{equation}
\begin{equation}
G^2_0 = - \left( \frac{2\dot{R}^2}{R^4} - \frac{\ddot{R}}{R^3} \right) (2 a \,
        \mathrm{sin} \, \theta) - \frac{2\dot{R}}{R^3 r} (a \, \mathrm{sin} 
        \, \theta)
\end{equation}
\begin{equation}
G^1_1 = - \left( \frac{\dot{R}^2}{R^4} - \frac{2\ddot{R}}{R^3} \right) (1 + 2 a r 
        \, \mathrm{cos} \, \theta) + \frac{4\dot{R}}{R^3} (a \, \mathrm{cos} \, 
        \theta) 
\end{equation}
\begin{equation}
G^1_2 = \frac{2\dot{R}}{R^3} (a r \, \mathrm{sin} \, \theta)
\end{equation}
\begin{equation}
G^2_1 = \frac{2\dot{R}}{R^3 r} (a \, \mathrm{sin} \, \theta)
\end{equation}
\begin{equation}
G^2_2 = G^3_3 = - \left( \frac{\dot{R}^2}{R^4} - \frac{2\ddot{R}}{R^3} \right) 
        (1 + 2 a r \, \mathrm{cos} \, \theta) + \frac{6\dot{R}}{R^3} (a \, 
        \mathrm{cos} \, \theta) 
. 
\end{equation}

The energy-momentum tensor appears to contain a null fluid with
\begin{equation}
{\tau l^0 l_0} = {\tau l^1 l_0} = - {\tau l^0 l_1} = -{\tau l^1 l_1} = 
- \frac{2\dot{R}}{\kappa R^3} (a \, \mathrm{cos} \, \theta)
,
\end{equation}
where the null vector is $l^a = (1,-1,0,0)$.
It is clear this does not fall off radially, so it is not 
consistent with radiation from $r=0$.
This suggests it is somehow being radiated by the fluid or is present at 
all times.

Looking at the remaining terms of the Einstein tensor with the null fluid 
component ${G^a_b}_{(nf)}$ subtracted reveals 
\begin{equation}
G^0_0 - {G^0_0}_{(nf)} = \frac{3\dot{R}^2}{R^4} (1 + 2 a r \, 
\mathrm{cos} \, 
        \theta) + \frac{6\dot{R}}{R^3} (a \, \mathrm{cos} \, \theta) 
\end{equation}
\begin{equation}
G^0_1 - {G^0_1}_{(nf)} = 0 
\end{equation}
\begin{equation}
G^1_0 - {G^1_0}_{(nf)} = \left( \frac{2\dot{R}^2}{R^4} - 
\frac{\ddot{R}}{R^3} 
        \right) (4a r \, \mathrm{cos} \, \theta)
\end{equation}
\begin{equation}
G^1_1 - {G^1_1}_{(nf)} = G^2_2 = G^3_3 = - \left( \frac{\dot{R}^2}{R^4} - 
        \frac{2\ddot{R}}{R^3} \right) (1 + 2 a r \, \mathrm{cos} \, 
        \theta) + \frac{6\dot{R}}{R^3} (a \, \mathrm{cos} \, \theta) 
.
\end{equation}
Due to axial symmetry, there must be no $u^3$ velocity field component, 
but since $G^1_1 - {G^1_1}_{(nf)} = G^2_2 = G^3_3$, it then makes it 
impossible to interpret the remaining mass-energy component as an isotropic 
perfect fluid with non-zero $u^1$ and $u^2$.
The energy-momentum tensor seems most logically interpreted with 
$u^1=u^2=u^3=0$, such that the fluid is an imperfect fluid that is 
comoving with the reference frame.
An imperfect fluid can be written as
\begin{equation}
T^a_b = (\mu +p)u^a u_b +pg^a_b + q^a u_b + u^a q_b + \pi^a_b
,
\end{equation}
where $q^a$ is the heat flow vector and $\pi^a_b$ is the anisotropic stress.
Thus, this appears to be an imperfect fluid with heat conduction 
components $q^1 u_0$, $u^0 q_2$, and $q^2 u_0$, and anisotropic stress 
components $\pi^1_2$ and $\pi^2_1$.
The anisotropic stress must obey $\pi^a_b u^b =0$, which with only $u^0$ 
non-zero means that $\pi^a_0 u^0 =0$, which is satisfied as long as all 
$\pi^a_0=0$, so the $T^a_0$ components of the energy-momentum tensor 
contain a contribution due to $q^a$ and not $\pi^a_b$.
The heat conduction obeys $q^a u_a =0$ and the anisotropic stress obeys 
$\pi^a_a=0$, so these do not show up in the trace of the Einstein tensor. 

Thus, it is possible to solve for the energy density and isotropic 
pressure via
\begin{equation}
G^a_a = \kappa (\mu - 3p) = \frac{6\ddot{R}}{R^3}(1+2ar \, \mathrm{cos} \, 
        \theta)+ \frac{24\dot{R}}{R^3}(a \, \mathrm{cos} \, \theta)
.
\end{equation}
There must be no $u^3$ velocity field component.
Assuming $\pi^3_3=0$, then $G^3_3 = - \kappa p$, which yields
\begin{equation}
p = \frac{1}{\kappa} \left( \frac{\dot{R}^2}{R^4} - \frac{2\ddot{R}}{R^3} \right) 
        (1 + 2 a r \, \mathrm{cos} \, \theta)  - \frac{6\dot{R}}{\kappa R^3} 
        (a \, \mathrm{cos} \, \theta) 
,
\end{equation}
and
\begin{equation}
\mu = \frac{3\dot{R}^2}{\kappa R^4} (1 + 2 a r \, \mathrm{cos} \, \theta) + 
        \frac{6\dot{R}}{\kappa R^3} (a \, \mathrm{cos} \, \theta)
.
\end{equation}

For positive expansion $\dot{R}/R$ and positive $a \, \mathrm{cos} \, \theta$, 
the energy density is always positive.  For negative $a \, 
\mathrm{cos} \, \theta$ the energy density can be negative.
At small times $\dot{R}^2/R^4 >> \dot{R}/R^3$, so the sign of the energy 
density depends on $1+2ar \, 
\mathrm{cos} \, \theta$, where $r \, \mathrm{cos} \, \theta$ is 
essentially an axial co-ordinate $z$ that specifies a plane in 
space; thus, the energy density is positive for $z > -1/(2a)$ at small 
times, so that in the limit as the acceleration goes to zero, the energy 
density is positive throughout the space.  
At late times $\dot{R}^2/R^4 << \dot{R}/R^3$, so the sign of the energy 
density depends on the sign of $a \, \mathrm{cos} \, \theta$, such that 
the energy density is only positive for $z > 0$.

At $\theta = \pi$, the pressure is positive for $R(t)=(t/t_0)^x$ 
with $0 < x \le 2$, the pressure is zero for $x = 2$, and the 
pressure is negative for $x > 2$.
At small times, the sign of the pressure reverses across a plane for $z 
= -1/(2a)$, but the magnitude of the pressure is no greater than 
$\mu/3$ as long as $x \ge 1$, so then the energy conditions are obeyed 
wherever the energy density is positive.
At late times, the sign of the pressure depends on the sign 
of $- a \, \mathrm{cos} \, \theta$, such that the pressure is positive for 
$z < 0$ and negative for $z > 0$, and the pressure approaches $p=-\mu$, 
violating the strong energy condition.

In the case of a radiation universe scale factor, $R(t) = t/t_0$, the 
pressure is 
\begin{equation}
p = \frac{1}{\kappa R^2 t^2} (1 + 2 a r \, \mathrm{cos} \, \theta)  - 
\frac{6}{\kappa R^2 t} (a \, \mathrm{cos} \, \theta) 
\end{equation}
and the energy density is
\begin{equation}
\mu = \frac{3}{\kappa R^2 t^2} (1 + 2 a r \, \mathrm{cos} \, \theta) + 
        \frac{6}{\kappa R^2 t} (a \, \mathrm{cos} \, \theta)
.
\end{equation}
Thus, at small times the pressure is negative for $z < -1/(2a)$ and 
positive for $z > -1/(2a)$, while at large times the pressure becomes 
negative for $z > 0$ and positive for $z < 0$.
The equation of state can be written as
\begin{equation}
p = \frac{\mu_r}{3} - \mu_{\Lambda}
,
\end{equation}
if the energy density is interpreted as having a radiation component $\mu_r$ given by
\begin{equation}
\mu_r = \frac{3}{\kappa R^2 t^2} (1 + 2 a r \, \mathrm{cos} \, \theta)
\end{equation}
and a component with $\Lambda$-type equation of state given by
\begin{equation}
\mu_{\Lambda} = \frac{6}{\kappa R^2 t} (a \, \mathrm{cos} \, \theta)
.
\end{equation}
It should be noted that this $\Lambda$-type component cannot be interpreted
geometrically as a cosmological constant since it is a function of $\theta$
and $t$.  It could be interpreted as an inhomogeneous dark energy, although
clearly it will yield a negative energy density for $z<0$, so it violates the
weak energy condition (and the dominant energy condition) and should be 
considered physically unrealistic for $z<0$, unless we are willing to admit 
exotic matter as being physically plausible.  
While the dominant and weak energy conditions are obeyed for $z>0$, the 
strong energy condition is violated for $z>0$, but these facts are 
universally true of dark energy that has positive energy density, so it 
is in no way physically different that the strong energy condition is 
violated.

In the case of a dust universe scale factor, $R(t) = (t/t_0)^2$,
the pressure is
\begin{equation}
p = - \frac{12}{\kappa R^2 t} (a \, \mathrm{cos} \, \theta) 
\end{equation}
and the energy density is
\begin{equation}
\mu = \frac{12}{\kappa R^2 t^2} (1 + 2 a r \, \mathrm{cos} \, \theta) +
        \frac{12}{\kappa R^2 t} (a \, \mathrm{cos} \, \theta)
.
\end{equation}
For all time the pressure is positive for $z < 0$ and negative for $z > 0$.
There is only one term in the pressure gradient, so this appears to be 
what is causing the imperfect fluid to accelerate with the reference 
frame, although the heat conduction may also be involved as well.
The equation of state can be written as
\begin{equation}
p = -\mu+\frac{12}{\kappa R^2 t^2} (1 + 2 a r \, \mathrm{cos} \, \theta)
,
\end{equation}
which is most easily explained if the energy density is assumed to have a 
pressureless dust component $\mu_d$ given by 
\begin{equation}
\mu_d = \frac{12}{\kappa R^2 t^2} (1 + 2 a r \, \mathrm{cos} \, \theta)
\end{equation}
and, as above, an inhomogeneous dark energy component $\mu_{\Lambda}$ given by
\begin{equation}
\mu_{\Lambda} = \frac{12}{\kappa R^2 t} (a \, \mathrm{cos} \, \theta)
.
\end{equation}

\section{A cosmological photon rocket}

Performing a conformal transformation on Kinnersley's photon rocket yields
\[
\mathrm{d}s^2 = [R(t)]^2 \Bigg( - \mathrm{d}{t}^2 + \mathrm{d}r^2 + r^2 
\left(\mathrm{d}{\theta}^2 
+ \mathrm{sin}^2 \theta \; \mathrm{d} {\phi}^2 \right) + 
2 a r^2 \mathrm{sin} \, \theta \; \mathrm{d}\theta (\mathrm{d}t-\mathrm{d}r) 
\]
\begin{equation}
+ \left(a^2 r^2 \mathrm{sin}^2 \theta + 2 a r \, \mathrm{cos} \, \theta + 
\frac{2m(t,r)}{r} \right)(\mathrm{d}t-\mathrm{d}r)^2 \Bigg)
.
\end{equation}
Looking at the Einstein tensor, the non-zero components are
\begin{equation}
G^0_0 = \frac{3\dot{R}^2}{R^4} \left(1 + 2 a r \, \mathrm{cos} \, \theta + 
        \frac{2m}{r} \right) + \frac{2\dot{R}}{R^3} \left(4 a \, \mathrm{cos} \, 
        \theta + \frac{2m}{r^2} - \frac{m'}{r} \right) + \frac{2}{R^2 r^2} 
        (3am \, \mathrm{cos} \, \theta - m')
\end{equation}
\begin{equation}
G^0_1 = \frac{2\dot{R}}{R^3} \left( - a \, \mathrm{cos} \, \theta + 
        \frac{m}{r^2} + \frac{m'}{r} \right) - \frac{2}{R^2 r^2} (3am \, 
        \mathrm{cos} \, \theta - m')
\end{equation}
\[
G^1_0 = \left( \frac{\dot{R}^2}{R^4} - \frac{2\ddot{R}}{R^3} \right) \left(4a 
        r \, \mathrm{cos} \, \theta + \frac{4m}{r} \right) + 
        \frac{2\dot{R}}{R^3} \left( a \, \mathrm{cos} \, \theta - 
        \frac{m}{r^2} - \frac{m'}{r} \right) 
\]
\begin{equation}
+ \frac{2}{R^2 r^2} (3am
        \, \mathrm{cos} \, \theta - m')
\end{equation}
\begin{equation}
G^0_2 = \frac{2\dot{R}}{R^3} (a r \, \mathrm{sin} \, \theta)
\end{equation}
\begin{equation}
G^2_0 = - \left( \frac{2\dot{R}^2}{R^4} - \frac{\ddot{R}}{R^3} \right) (2 a \,
        \mathrm{sin} \, \theta) - \frac{2\dot{R}}{R^3 r} \left( 1 + 
        \frac{3m}{r} \right) a \, \mathrm{sin} \, \theta
\end{equation}
\[
G^1_1 = - \left( \frac{\dot{R}^2}{R^4} - \frac{2\ddot{R}}{R^3} \right) \left(1 
        + 2 a r \, \mathrm{cos} \, \theta + \frac{2m}{r} \right) + 
        \frac{2\dot{R}}{R^3} \left( 2a \, \mathrm{cos} \, \theta + 
        \frac{4m}{r^2} + \frac{m'}{r} \right) 
\]
\begin{equation}
- \frac{2}{R^2 r^2} (3am
        \, \mathrm{cos} \, \theta - m')
\end{equation}
\begin{equation}
G^1_2 = \frac{2\dot{R}}{R^3} (a r \, \mathrm{sin} \, \theta)
\end{equation}
\begin{equation}
G^2_1 = \frac{2\dot{R}}{R^3 r} \left( 1 + \frac{3m}{r} \right) a \, 
        \mathrm{sin} \, \theta
\end{equation}
\begin{equation}
G^2_2 = G^3_3 = - \left( \frac{\dot{R}^2}{R^4} - \frac{2\ddot{R}}{R^3} \right) 
        \left(1 + 2 a r \, \mathrm{cos} \, \theta + \frac{2m}{r} \right) + 
        \frac{6\dot{R}}{R^3} (a \, \mathrm{cos} \, \theta) 
. 
\end{equation}
The only terms that were not present in the $a=0$ or $m=0$ cases of the 
previous two sections are the $6am \, \mathrm{cos} \, \theta / (R^2 r)$ 
terms in $G^0_0$, $G^0_1$, $G^1_0$, and $G^1_1$, and the $6\dot{R}am \, 
\mathrm{sin} \, \theta / (R^2 r^2)$ terms in $G^2_0$ and $G^2_1$. 
The former are simply the anisotropic 
null fluid terms from Kinnersley's photon rocket scaled down by $R^2$, 
and the latter appear to be in the heat conduction and anisotropic stress 
terms of the imperfect fluid.

The null fluid is interpreted as
\begin{equation}
{\tau l^0 l_0} = {\tau l^1 l_0} = - {\tau l^0 l_1} = -{\tau l^1 l_1} = 
        - \frac{2\dot{R}}{\kappa R^3} \left( a \, \mathrm{cos} \, \theta - 
        \frac{m}{r^2} - \frac{m'}{r} \right) - \frac{2}{\kappa R^2 r^2} 
        (3am \, \mathrm{cos} \, \theta - m')
.
\end{equation}

Looking at the remaining terms of the Einstein tensor with the null fluid 
component ${G^a_b}_{(nf)}$ subtracted reveals 
\begin{equation}
G^0_0 - {G^0_0}_{(nf)}= \frac{3\dot{R}^2}{R^4} \left(1 + 2 a r \, 
\mathrm{cos} \, \theta + 
        \frac{2m}{r} \right) + \frac{6\dot{R}}{R^3} \left( a \, 
\mathrm{cos} \, \theta + \frac{m}{r^2} \right) 
\end{equation}
\begin{equation}
G^0_1 -{G^0_1}_{(nf)} = 0
\end{equation}
\begin{equation}
G^1_0 - {G^1_0}_{(nf)} = 4 \left( \frac{\dot{R}^2}{R^4} - 
        \frac{2\ddot{R}}{R^3} \right) \left(a 
        r \, \mathrm{cos} \, \theta + \frac{m}{r} \right)
\end{equation}
\begin{equation}
G^1_1 - {G^1_1}_{(nf)} = - \left( \frac{\dot{R}^2}{R^4} - 
        \frac{2\ddot{R}}{R^3} \right) \left(1 
        + 2 a r \, \mathrm{cos} \, \theta + \frac{2m}{r} \right) + 
        \frac{6\dot{R}}{R^3} \left( a \, \mathrm{cos} \, \theta + 
        \frac{m}{r^2} \right) 
.
\end{equation}
Since $u^3$ must be zero due to axial symmetry, and $G^2_2 = G^3_3$, it 
appears that $u^2$ is once again zero as with the case of the 
accelerating cosmological reference frame, while $G^1_1 \neq G^3_3$ 
suggests that $u^1$ is once again non-zero as with the case of the 
cosmological radiating stellar exterior.

Again interpreting the remainder of the energy-momentum tensor as an  
imperfect fluid yields 
\begin{equation}
G^a_a = \kappa (\mu - 3p) = \frac{6\ddot{R}}{R^3} \left( 1+2ar \, \mathrm{cos} \, 
        \theta + \frac{2m}{r} \right) + \frac{12\dot{R}}{R^3} \left(2a \,  
        \mathrm{cos} \, \theta + \frac{m}{r^2} \right) 
. 
\end{equation}
Since $u^3$ must be zero, the pressure is interpreted as $G^3_3=-\kappa 
p$ so that 
\begin{equation}
p = \frac{1}{\kappa} \left( \frac{\dot{R}^2}{R^4} - \frac{2\ddot{R}}{R^3} \right) 
        \left(1 + 2 a r \, \mathrm{cos} \, \theta + \frac{2m}{r} \right) - 
        \frac{6\dot{R}}{\kappa R^3} (a \, \mathrm{cos} \, \theta) 
,
\end{equation}
which means that the energy density is given by
\begin{equation}
\mu = \frac{3\dot{R}^2}{\kappa R^4} \left( 1 + 2 a r \, \mathrm{cos} \, \theta 
      + \frac{2m}{r} \right) + \frac{6\dot{R}}{\kappa R^3} \left(a \, 
      \mathrm{cos} \, \theta +\frac{2m}{r^2} \right)
.
\end{equation}

In the case of the accelerating cosmological reference frame of the 
previous section, the anisotropic stress obeys $\pi^a_0 u^0 =0$, such that 
all $\pi^a_0$ have to be zero.
In the case of the rocket with a radial velocity component $u^1$, the 
anisotropic stress must obey $\pi^a_0 u^0 + 
\pi^a_1 u^1 = 0$, so with $\pi^2_1 \neq 0$, it requires $\pi^2_0 \neq 
0$, so the new terms $6\dot{R}am \, \mathrm{sin} \, \theta / (R^2
r^2)$ in $G^2_0$ and $G^2_1$ appear to both be anisotropic stress terms in 
the imperfect fluid that result due to the mass of the rocket.

The additional terms in the energy density due to the mass of the rocket 
act to keep $\mu$ positive at small $r$, but part of $\mu$ is  
negative and causes it to violate the weak and dominant energy conditions 
for negative $a \, \mathrm{cos} \, \theta$.
As with the $m=0$ case, all energy conditions are satisfied for positive 
$a \, \mathrm{cos} \, \theta$, other than the strong energy condition 
that is violated at late times due to the dominance of the $\Lambda$-type 
component of the fluid.

\section{Summary and discussion}

A physical solution of Einstein's field equations has been found for Vaidya's 
radiating stellar exterior in a cosmological background that is a 
two-fluid consisting of a null fluid and a cosmological perfect fluid.  
In the limit of infinite radius, the density and pressure are the same as 
the standard cosmological solutions. 

It is interesting that cosmological white holes seem to more readily 
yield physical solutions than cosmological black holes do.
Perhaps this is due to the tendency of black holes to oppose the expansion of 
the universe, since the expansion of space can cause outgoing null 
geodesics to expand instead of being trapped.
For shrinking universes the sign of $\dot{R}/R$ is reversed, 
changing the energy density such that the cosmological white holes 
are unphysical and the cosmological black holes are physical.
Thus, there is no physical preference for white hole solutions over 
black hole solutions, unless the universes of consideration are restricted
to be expanding.

The spacetimes for an accelerating cosmological reference frame and 
Kinnersley's photon rocket in cosmological background have been 
interpreted as two-fluid models consisting of a null fluid and an imperfect 
fluid.
On the leading side of the rocket, only the strong energy condition is 
violated at late times (as is always the case for a lambda-type equation
of state with positive energy density), while on the trailing side the 
region satisfying the weak and dominant energy conditions is inversely 
related to the acceleration and the entire region on the trailing side of 
the rocket becomes unphysical at late times.
Since acceleration is inversely related to the inertial mass $(\mu+p)$, 
it appears that in order to maintain the acceleration the 
energy density has to be negative on the trailing side of the rocket in 
order to offset the increasing pressure required to maintain the pressure 
gradient.
If the energy density is interpreted as arising due to separate components,
one of them being an inhomogeneous dark energy component, then this dark 
energy component has negative energy density on the trailing side
of the rocket, so this interpretation requires the entire trailing side of 
the rocket to violate the weak energy condition and be unphysical for any 
finite acceleration.

While it may not be any more realistic to conceive of a universe that 
accelerates in bulk than one that undergoes bulk rotation, spacetimes 
for universes where the entire matter distribution undergoes acceleration 
serve to show that acceleration is absolute in general relativity and not 
simply relative.  In the cosmological photon rocket spacetime, an observer 
will have empirical evidence that acceleration is taking place: without even 
observing the unusual mass-energy distribution of the universe and inferring 
that the matter is accelerating, an observer subject to the pressure gradient
must directly ``feel'' the acceleration due to the unbalanced forces that
act on the observer.  This differs from spacetimes with rotation such as van 
Stockum dust (Lanczos \cite{Lan24}) and the G\"{o}del universe \cite{God49} 
where the matter is simply moving inertially and there is no such empirically 
apparent pressure gradient or other force that acts to accelerate the matter.

\begin{acknowledgments}

Thanks go to Charles Hellaby and Andrzej Krasi\'{n}ski for useful discussions 
that precipitated this work.  Additional thanks go to Charles Hellaby for 
providing many useful comments on this paper.  This research was funded by 
the National Research Foundation of South Africa.

\end{acknowledgments}


\begin{thebibliography}{99}

\bibitem{Tha81} S.~N.~G.~Thakurta, Indian J.\ Phys.\ \textbf{55B}, 304 
(1981).
\bibitem{Kra97} A.~Krasi\'{n}ski, \textit{Inhomogeneous Cosmological 
Models} (Cambridge U. Press, Cambridge, 1997) p.~255.
\bibitem{McC06} M.~L.~McClure and C.~C.~Dyer, Class.\ Quantum Grav.\ 
\textbf{23}, 1971 (2006).
\bibitem{Sul05} J.~Sultana and C.~C.~Dyer, Gen.\ Relativ.\ Gravit. 
\textbf{37}, 1349 (2005).
\bibitem{Ker65} R.~P.~Kerr and A.~Schild, in \textit{Atti del Convegno 
Sulla Relativita Generale: Problemi Dell'Energia E Onde Gravitazionale}, 
edited by G.~Barbera (Comitato Nazionale per le Manifestazione 
celebrative, Florence, 1965) p.~222.
\bibitem{Sul03} J.~Sultana, Ph.D.\ thesis, University of Toronto, 2003.
\bibitem{Vai43} P.~C.~Vaidya, Current Sci.\ \textbf{12}, 
183 (1943) [Gen.\ Relativ.\ Gravit.\ \textbf{31}, 119 (1999)]. 
\bibitem{Vai73} P.~C.~Vaidya and L.~K.~Patel, Phys.\ Rev.\ D \textbf{7}, 
3590 (1973).
\bibitem{Kin69} W.~Kinnersley, Phys.\ Rev.\ \textbf{186}, 1335 (1969). 
\bibitem{Kin70} W.~Kinnersley and M.~Walker, Phys.\ Rev. D 
\textbf{2}, 1359 (1970).
\bibitem{Har94} J.~H.~Harper and C.~C.~Dyer, \textit{Tensor Algebra 
with REDTEN} \\ 
(http://www.scar.utoronto.ca/$\sim$harper/redten/root.html, 1994).
\bibitem{Vai89} P.~C.~Vaidya and L.~K.~Patel, Pramana---J.\ Phys.\ 
\textbf{32}, 731 (1989).
\bibitem{Bon94} W.~B.~Bonnor, Class.\ Quantum Grav.\ \textbf{11}, 2007 
(1994).
\bibitem{Wal84} R.~M.~Wald, \textit{General Relativity} (U. of Chicago 
Press, Chicago, 1984) p.~219. 
\bibitem{Lan24} K. Lanczos. Z.\ Phys.\ \textbf{21}, 73 (1924) 
[Gen.\ Relativ.\ Gravit. \textbf{29}, 363 (1997)].
\bibitem{God49} K. G\"{o}del, Rev.\ Mod.\ Phys.\ \textbf{21}, 447 (1949).

\end{thebibliography}
\end{document}